\documentstyle[prl,aps,epsfig
,twocolumn
]{revtex}
\begin{document}
\draft

\twocolumn[
\hsize\textwidth\columnwidth\hsize\csname @twocolumnfalse\endcsname

\title{Quantum Monte Carlo calculation of the finite
 temperature Mott-Hubbard transition}

\author{Jaewook Joo and Viktor Oudovenko}

\address{Serin Physics Laboratory, Rutgers University,
136 Frelinghuysen Road, Piscataway, New Jersey 08854, USA}
\date{\today}

\maketitle

\begin{abstract}
We present clear numerical evidence for the coexistence of metallic and
insulating dynamical mean field theory(DMFT) solutions in a half-filled
single-band Hubbard model with bare semicircular density of states at
finite temperatures. Quantum Monte Carlo(QMC) method is used to
solve the DMFT equations. We discuss important technical aspects  of
the DMFT-QMC which need to be taken into account in order to obtain the
reliable results near the coexistence region. Among them are the
critical slowing down of the iterative solutions  near phase
boundaries, the convergence criteria for the DMFT iterations, the
interpolation of the discretized Green's function and the reduction of
QMC statistical and systematic errors. Comparison of our results with
those of other numerical methods is presented in a phase diagram.
\end{abstract}


]
\narrowtext

The Mott-Hubbard Transition~\cite{Mott:1990,Hubbard:1964} is an
outstanding open problem in condensed matter physics. The theoretical
progress in this field has been impeded by non-perturbative nature of
the strongly correlated system.

In recent years considerable progress has been made with a single band
Hubbard model by using DMFT approximation which becomes exact in the
limit of infinite coordinations ($z=\infty$)~\cite{Review:1996}. The
most striking point of these works is a discontinuous metal-insulator
transition(MIT) at finite temperature $T$ which takes place at the
critical interaction strength line $ U_{c}(T)$. This line lies inside a
coexistence region  where two phases, one metallic and one insulating
coexist. The coexistence region is bounded by two lines $U_{c1}(T)$ and
$U_{c2}(T)$ and $U_{c1}(T)< U_{c}(T)< U_{c2}(T)$ \cite{Review:1996}.

While this picture is now generally accepted \cite{dieter} by all
groups working on the problem, the origin of the controversial results
over the coexistence region in a single-band Hubbard model at finite
temperature within QMC method, and more generally the difficulties in
carrying out the DMFT-QMC algorithm
~\cite{Schlipf:1999,Rozenberg:1999,Krauth:1999}, have not been
clarified. This is an important problem, since DMFT-QMC is currently
being applied to many similar models with more complicated orbital and
spin structure.

In this paper we provide clear evidence for the coexistence of metallic
and insulating solutions in a half-filled single-band Hubbard model
with bare semicircular density of states at finite temperature within
DMFT-QMC calculation~\cite{QMC}. We discuss several  aspects of
DMFT-QMC method  which have to  be taken into account in order to get
the reliable results near a coexistence region. First, we identify the
source of the main difficulties in running DMFT-QMC near the
coexistence region: uncontrolled errors in DMFT-QMC in the region where
critical slowing down is dominant. The critical slowing down of the
DMFT iteration follows from the Landau analysis \cite{Kotliar:2000} and
can only be avoided by  choosing a different rearrangement of the DMFT
equation such as the one suggested by the Newton method. In this paper,
however we use this  critical slowing down as an indicator which allows
us to locate boundaries of the coexistence region, and to fully confirm
,using QMC methods, the predictions of the Landau analysis. We discuss
how to  minimize errors in DMFT-QMC calculation by using correct
interpolation of the discretized Green's functions in direct Fourier
Transformation($FT$) as well as a proper treatment of discontinuities
of Green's functions in inverse Fourier Transformation($iFT$). Also
extrapolation of data to the limit of $\delta
\tau(=\beta/L)\rightarrow0$ is used to reduce Trotter errors. Finally
we obtain  reliable coexistent insulating and metallic solutions within
DMFT-QMC method and estimate the location of the first order MIT line
$U_{c}(T)$.

In the limit of infinite dimensions a single-band Hubbard model
is mapped onto an Anderson impurity model supplemented by
a self-consistency condition~\cite{Review:1996}.
DMFT equation of this model,

\begin{equation}
t^{2}G(i\omega)[\Delta]=\Delta(i\omega),
\end{equation}

\noindent
is solved iteratively within QMC calculation, where $\Delta(i\omega)$
and $G(i\omega)[\Delta]$ are a hybridization function and a local
Green's function of the impurity model. Our energy unit is set to be
$t=1/2$ which comes from a scaled hopping amplitude, $t/\sqrt{z}$, in
an infinite dimensional single-band Hubbard model.

Within DMFT-QMC calculation, the coexistence region should be explored
with a great care due to the following two reasons: critical slowing
down and errors in DMFT-QMC. The critical slowing down is the critical
phenomena which manifests itself by slowing down of the convergence
rate of a numerical method. The critical slowing down at a phase
boundary can be proved with the argument based on Landau free
energy~\cite{Kotliar:2000}: no convergence at an inflection point of
Landau free energy within iteration method. At a boundary of the
coexistence region one solution never converges while another one
converges fast. However around Mott endpoint  (where $U_{c1}$ and
$U_{c2}$ merge) both solutions converge slowly because the narrow
coexistence region around Mott endpoint brings two boundaries close
enough so as to make a whole coexistence region numerically unstable as
was pointed out in ref.~\cite{Kotliar:2000}. Unless carefully taken
care of, fluctuation of a solution originating from errors in DMFT-QMC
may result in undesirable transition of one solution to another one in
DMFT iteration around Mott endpoint where critical slowing down is
dominant.

In this paper we show that one can use the existence of critical
slowing down as an indicator of boundaries of the coexistence region.
Near a coexistence boundary, $U_{c1}(T)$ or $U_{c2}(T)$, metallic and
insulating solutions show different convergence patterns within DMFT
iterations: one converges very slowly while the other converges in a
few iterations. A convergence criterion is applied to determine
required number of iterations to get a convergent solution. We locate
the coexistence boundaries at which such iteration numbers diverge.

\begin{figure}
\begin{center}
\includegraphics[height=6cm,angle=0]{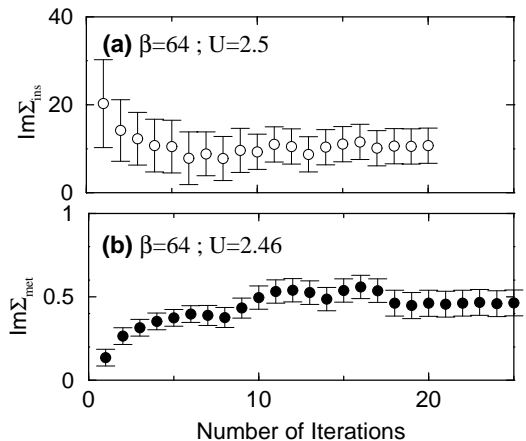}
\caption{ Convergence criterion. The error bars are the QMC fluctuations
at individual iterations. (a)Insulating solution. (b) Metallic solution.
}
\label{fig:1}
\end{center}
\end{figure}

The convergence of a solution is determined by monitoring of the most
fluctuating component $Im\Sigma(i \pi T)$, imaginary part of self
energy at the first Matsubara frequency, as a function of number of
iterations. QMC statistical error of $Im\Sigma(i \pi T)$ at each
iteration is controlled  by the  number of QMC sweeps while the change
in $Im\Sigma(i \pi T)$ between iterations is controlled by the  number
of iterations. The iteration procedure is  stopped when the change in
$Im\Sigma(i \pi T)$ from iteration to iteration is less than the QMC
statistical fluctuation of $Im\Sigma(i \pi T)$, $\delta_{QMC}
Im\Sigma\simeq\ \delta ImG/(ImG)^{2}$. Notice that the QMC fluctuation
of insulating solution is five to twenty times larger than QMC
fluctuation of metallic solution as in Fig.~\ref{fig:1} mostly because
$ImG_{ins}\ll ImG_{met}$. This convergence criterion is more strict
than the conventional one, which requires to stop iteration procedure
when the norm of $||Im\Sigma^{(i+1)}-Im\Sigma^{(i)}||$ is smaller than
a tolerance. For the conventional criterion underestimates the
contribution of the most fluctuating component to the convergence by
taking average of self energy differences between iterations over whole
frequency range. The convergence criteria are illustrated in
Fig.~\ref{fig:1}.

With this convergence criterion we performed the DMFT-QMC calculation
to find solutions of a half-filled single-band Hubbard model at
temperatures $T$=1/57, 1/64, 1/100, which turn out to be well below the
critical temperature $T_{c}$. A careful selection of initial seeding to
feed DMFT iteration is an essential part of finding of the coexistence
at finite temperature. At a given temperature, the best educated
seeding is used to start DMFT iteration at initial value of interaction
strength $U$: small interaction strength $U$ for metallic solution and
large $U$ for insulating solution. After DMFT-QMC run we obtain
convergent metallic and insulating solutions within a few iterations.
We increase interaction strength $U$ by small increment(0.01) for
metallic solution and decrease $U$ for insulating one. To reduce
computational time, the convergent solution at the previous value of
interaction strength $U$ is used as initial seeding for DMFT-QMC run at
the next value of interaction strength $U$. We observe

\begin{figure}
\begin{center}
\includegraphics[height=6cm,angle=0]{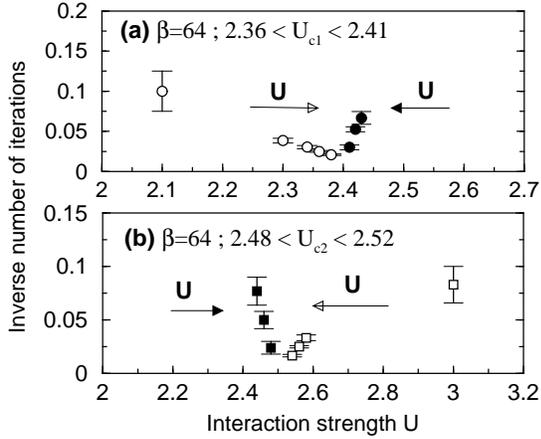}
\caption{ The critical slowing down at boundaries of the coexistence
region, $U_{c1}(T)$ and $U_{c2}(T)$. Solid(or open) symbols are
obtained by approaching $U_{c1}$ and $U_{c2}$ from inside(or outside)
the coexistence region along solid(or open) arrows. } 
\label{fig:2}
\end{center}
\end{figure}

\noindent
convergence pattern of solutions and repeat the procedures
until critical slowing down is noticeably developed. By plotting
inverse number of required iterations for the convergence of insulating
solution in Fig.~\ref{fig:2}(a) and of metallic solution in
Fig.~\ref{fig:2}(b), we identify upper bound of $U_{c1}(T)$ and lower
bound of $U_{c2}(T)$. To find lower bound of $U_{c1}(T)$ and upper
bound of $U_{c2}(T)$ we approach coexistence boundaries, $U_{c1}(T)$
and $U_{c2}(T)$, from outside the coexistence region. Outside the
coexistence region only one solution exists. For interaction strength
$U(T)<U_{c1}(T)$ DMFT iteration which starts with insulating seeding
would produce convergent metallic solution. As $U$ gets closer to
$U_{c1}$ from outside the coexistence region, convergence rate become
slower. Such critical slowing down is clearly captured with open
symbols in Fig.~\ref{fig:2}(a). By approaching coexistence boundaries
from both inside and outside the coexistence region we obtain upper and
lower bounds of $U_{c1}$ and $U_{c2}$ at several temperatures: for
$\beta=57$, $2.36\leq U_{c1}(T)\leq 2.4$ and $2.42\leq U_{c2}(T)\leq
2.53$; for $\beta=64$, $2.36\leq U_{c1}(T)\leq 2.41$ and $2.48\leq
U_{c2}(T)\leq 2.52$; for $\beta=100$, $2.35\leq U_{c1}(T)\leq 2.45$ and
$2.58\leq U_{c2}(T)\leq 2.62$.

At $\beta=57$ the coexistence region is narrow enough so as to bring
the critical slowing down into play over the whole coexistence region.
Thus at temperatures above $\beta=57$ the coexistence region could
exist as well but will be inapproachable numerically for it is hard to
achieve convergent solutions.

All known errors in DMFT-QMC should be reduced as much as possible in
order to obtain irrefutable evidence of the coexistence. Those errors
are the QMC statistical errors and the systematic errors. The latter
originate from two sources: the finite sampling size errors in both
direct $FT$ and $iFT$ and the Trotter finite size errors.

To reduce the finite sampling size errors in direct $FT$ we use cubic
spline interpolation between grid points of Green's functions. The main
idea of cubic spline interp-

\noindent
olation is to get an interpolating function
that has continuous first and second derivatives at grid points.
Boundary conditions are necessary to close the set of liner equations
to determine interpolation function. The usual way, when boundary
conditions are unknown, is to put the second derivatives at end points
to zero (natural spline). In the system we treat we can obtain boundary
conditions from Green's function. To do so, we expand Green's function 
$G(i\omega)$ via k-th derivatives of $G(\tau)$ at end points:
$G(i\omega)$=$\sum_{k=0}^{\infty}{(-1)^{k+1}(G^{(k)}(\beta^{-})+G^{(k)}
(0^{+}))/(i\omega_{n})^{k+1}}$. One can also obtain correct high
frequency behavior of Green's functions using the moments of the
spectral density~\cite{Nolting:1997}. Making
correspondence between these two expansions we get the following
formula for the second derivatives of Green's functions in the case
of half filled Hubbard model within DMFT:
$G^{(2)}(0^{+})+G^{(2)}(\beta^{-})=-(U^{2}/4+t^{2})$ for local Green's
function and $G^{(2)}_{o}(0^{+})+G^{(2)}_{o}(\beta^{-})=-t^{2}$ for
Weiss function. The symmetry property of Green's function at half
filling is taken into account:
$G^{(k)}(0^{+})=(-1)^{k}G^{(k)}(\beta^{-1})$.

Despite of a lot of frequency points used to define Green's function
$G(i\omega)$, the finite frequency cutoff in $iFT$ results in
removal of the discontinuity of $G(\tau)$ at end points.
To correct this situation we should extract $1/i\omega$ tail from Green's
function before making $iFT$.
To get correct $iFT$ of $G(i\omega)$ one should make numerical $iFT$ for
$G(i\omega)-1/i\omega$ function and analytical $iFT$ for the tail $1/i\omega$.
In the case of half-filling the correct $iFT$ is:
$G(\tau)=T\sum_{n}e^{-i\omega_{n}\tau}[G(i\omega_{n})-\frac{1}{i\omega}]-1/2$
for $\tau> 0$.

The corrections in direct $FT$ and $iFT$ are implemented in DMFT-QMC:
correct boundary conditions in spline interpolation and tail correction
in $iFT$. All artificial oscillations at both Im$\Sigma(i\omega)$ and
$G(\tau)$ presented in DMFT-QMC without corrections disappears in
DMFT-QMC with corrections.

\begin{figure}
\begin{center}
\includegraphics[height=6cm, angle=0]{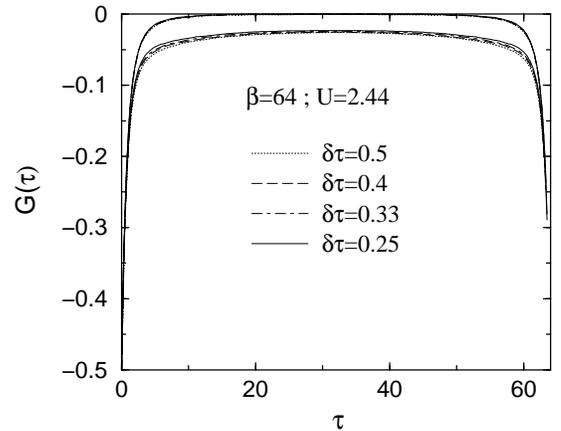}
\caption{ Convergence of local Green's function $G(\tau)$ in the limit
of $\delta\tau(=\frac{\beta}{L})\rightarrow0$. Upper(or lower) set of
lines correspond to insulating(or metallic) solutions. } \label{fig:3}
\end{center}
\end{figure}

\begin{figure}
\begin{center}
\includegraphics[height=6cm,angle=0]{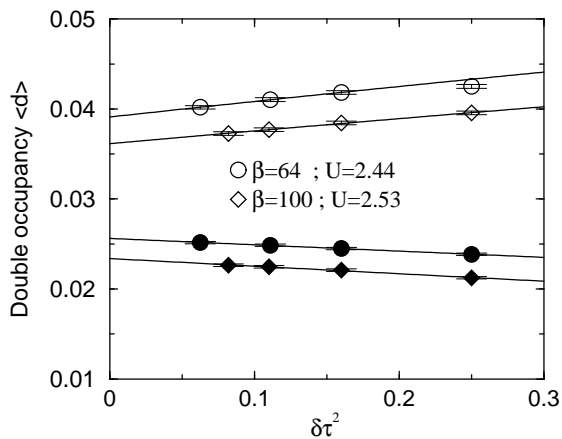}
\caption{ Convergence of double occupancy as a function of
$\delta\tau^{2}$. Open(or solid) symbols with QMC error bars correspond
to metallic(or insulating) double occupancies.
%Function $a\pm\sqrt{b+c\delta\tau^{2}}$ is used to fit numerical data.
}
\label{fig:4}
\end{center}
\end{figure}

Having implemented the corrections in DMFT-QMC, we perform
DMFT-QMC run to get the coexistence at numerically most stable points,
i.e., at mid-points between the coexistence boundaries. The
statistical errors are reduced by increase of QMC sweeps per site upto
$3\times10^{5}$. Several DMFT-QMC calculations are performed with
different values of $\delta\tau(=\frac{\beta}{L})$. Trotter errors are
removed by extrapolation of data to the limit of $\delta
\tau\rightarrow0$. In Fig.~\ref{fig:3} it is shown convergence of local
Green's function $G(\tau)$ in the limit of $\delta\tau\rightarrow0$.
Insulating solutions are all within the QMC fluctuation
$(\sim10^{-4})$. However metallic solutions show monotonous increases
with decreasing of $\delta\tau$. In Fig.~\ref{fig:4} it is shown
convergence of double occupancy, $\langle d \rangle =\langle
n_{\downarrow}n_{\uparrow} \rangle $, as a function of $\delta
\tau^{2}$. The convergence of metallic and insulating double
occupancies in the limit of $\delta\tau\rightarrow0$ confirms the
coexistence within DMFT-QMC. In the limit of $\delta\tau\rightarrow0$
we obtain $\langle d_{met} \rangle \simeq 0.039$ and $\langle d_{ins}
\rangle \simeq 0.026$ for $\beta=64$ and $U=2.44$; $\langle d_{met}
\rangle \simeq 0.036$ and $\langle d_{ins} \rangle \simeq 0.023$ for
$\beta=100$ and $U=2.53$. With reducing all other finite size errors
except ones originating from Trotter approximation, the effect of
Trotter errors on the convergence of solution is found to be small
compared to errors in direct $FT$ and $iFT$.

\begin{figure}
\begin{center}
\includegraphics[height=6cm,angle=0]{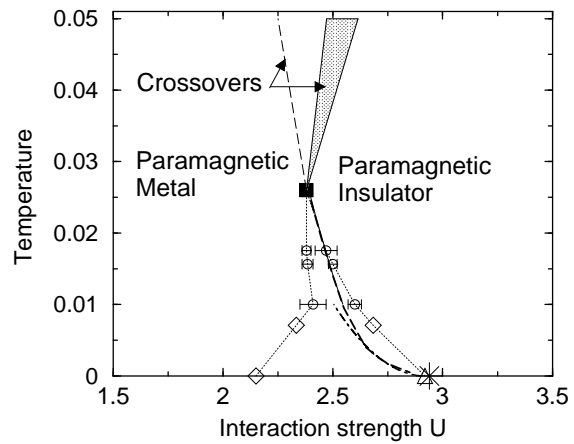}
\caption {Phase diagram of the paramagnetic metal to insulator
transition. Coexistence boundaries(circles)
with their uncertainties(error bars) within QMC.
MIT line (solid) from the Landau
analysis of IPT data. MIT line(dot-dashed) from low temperature
considerations(see text). The following data are from other numerical
works(see text for references). Second order critical point(square)
within QMC. $U_{c1}$ and $U_{c2}$ points(diamonds) 
and MIT line(thick long-dashed) within ED method.
T=0 MIT point from projective method(triangle) and from NRG(star).
Two crossovers(long-dashed line and shaded
area) above $T_{c}$. Coexistence boundaries from QMC and ED methods are
connected for guides to eye(dotted line).
}
\label{fig:5}
\end{center}
\end{figure}

We now use our QMC results together with previous  numerical results
obtained with other methods to draw the  phase diagram in
Fig.~\ref{fig:5}. To supplement our QMC data, we use the estimates of
the Mott endpoint within QMC~\cite{Rozenberg:1999}. We also include
estimates for the coexistence boundaries
~\cite{Laloux:1994,Rozenberg:1996} from Exact Diagonalization(ED). At
zero temperature, the value of $U_{c2}$ was obtained with great
accuracy in ref \cite{fisher} using the projective self consistent
method. Other exact methods such as the numerical renormalization group
method(NRG)~\cite{bulla} give a very similar value for this quantity. The
region near the Mott endpoint cannot be studied  with numerically exact
methods for the reasons discussed in this paper. However  we can use
the Landau theory to scale the results of the IPT near the critical
region as described in ref \cite{Kotliar:2000}. This approach was
used to determine the $U_{c1}(T)$ and $U_{c2}(T)$

\noindent
lines near the Mott endpoint as well as the location of the first order 
MIT line ${U_c}(T)$. Two crossover lines above the Mott endpoint
\cite{Rozenberg:1995} were scaled in a similar manner.
The physical meaning of the crossovers were
discussed in ref ~\cite{Rozenberg:1995} in connection with experimental
observations in $V_2 O_3$. Landau MIT line and crossover
lines are mapped onto QMC phase diagram by shifting of $U_{c}(T_{c})$ and
using the temperature scaling discussed in ref \cite{Kotliar:2000}.

Near zero temperature, we determine the location of the finite
temperature MIT line, by following ref \cite{fisher} and assuming an
entropy difference of $ln(2)$ between the metallic and the insulatining
phase. The MIT lines obtained from those two methods agree well with
MIT line obtained by ED method \cite{Moeller:1996} which we also
included in Fig.~\ref{fig:5}. Given the systematic finite size errors
of the various numerical approximations, it is clear that there is
quantitative consensus ( at the level of ten to twenty  percent)
between numerical calculations on where the coexistence region of fully
frustrated   Hubbard model in infinite dimensions lies. We note in
passing, that the line of second order phase transitions  reported in
ref.~\cite{Schlipf:1999}, is actually crossover line at high
temperature and the  $U_{c2}(T)$ line at low temperature.

In conclusion, we find that  QMC can be used to establish rigorously
the existence of a coexistence region in the frustrated Hubbard model
in infinite dimensions. Furthermore the numerical results are in full
agreement with the predictions of the Landau theory of
ref.~\cite{Kotliar:2000} below $T_c$. Finally we notice that the
methodology described in this paper should be useful when applying DMFT
to other phase transitions in strongly correlated systems.

We thank G. Kotliar for many useful and fruitful discussions. This
research was motivated by the stimulating discussions at the workshop
on Theoretical Methods for Strongly Correlated Fermions at the CRM in
Montreal. We have used the supercomputer at the Center for Advanced
Information Processing in Rutgers.

\end{document}